\documentstyle[12pt]{article}
\newlength{\extralineskip}
\newcommand{\beq}{\begin{equation}}
\newcommand{\eeq}{\end{equation}}

\newcommand{\bd}{\begin{displaymath}}
\newcommand{\ed}{\end{displaymath}}

\def\e{\, {\rm e}}
\def\IR{{\rm I}\!{\rm R}}
\def\inbar{\,\vrule height1.5ex width.4pt depth0pt}
\def\IC{\relax\hbox{$\inbar\kern-.3em{\rm C}$}}
\def\IR{\relax{\rm I\kern-.18em R}}
\def\IZ{\relax\ifmmode\mathchoice
{\hbox{\kern-.4em Z}}{\hbox{Z\kern-.4em Z}}
{\lower.9pt\hbox{Z\kern-.4em Z}}
{\lower1.2pt\hbox{Z\kern-.4em Z}}\else{Z\kern-.4em Z}\fi}

\def\ps2{{\bar{\psi}\psi}}
\makeatletter
\newdimen\normalarrayskip              
\newdimen\minarrayskip                 
\normalarrayskip\baselineskip
\minarrayskip\jot
\newif\ifold             \oldtrue            \def\new{\oldfalse}
\def\arraymode{\ifold\relax\else\displaystyle\fi} 
\def\@arrayskip{\ifold\baselineskip\z@\lineskip\z@
     \else
     \baselineskip\minarrayskip\lineskip2\minarrayskip\fi}
\def\@arrayclassz{\ifcase \@lastchclass \@acolampacol \or
\@ampacol \or \or \or \@addamp \or
   \@acolampacol \or \@firstampfalse \@acol \fi
\edef\@preamble{\@preamble
  \ifcase \@chnum
     \hfil$\relax\arraymode\@sharp$\hfil
     \or $\relax\arraymode\@sharp$\hfil
     \or \hfil$\relax\arraymode\@sharp$\fi}}
\def\@array[#1]#2{\setbox\@arstrutbox=\hbox{\vrule
     height\arraystretch \ht\strutbox
     depth\arraystretch \dp\strutbox
     width\z@}\@mkpream{#2}\edef\@preamble{\halign \noexpand\@halignto
\bgroup \tabskip\z@ \@arstrut \@preamble \tabskip\z@ \cr}%
\let\@startpbox\@@startpbox \let\@endpbox\@@endpbox
  \if #1t\vtop \else \if#1b\vbox \else \vcenter \fi\fi
  \bgroup \let\par\relax
  \let\@sharp##\let\protect\relax
  \@arrayskip\@preamble}
\makeatother

\begin{document}

\begin{titlepage}

\baselineskip=12pt

\rightline{UBC/S-96/1}
\rightline{OUTP-96-04P}
\rightline{hep-th/9602007}
\rightline{   }
\rightline{Revised Version}
\rightline{   }
\rightline{\today}

\vskip 0.5truein
\begin{center}

{\Large\bf Polymer Statistics and Fermionic Vector Models\footnote{This work
was supported in part by the Natural Sciences and Engineering Research Council
of Canada.}}\\
\vskip 0.5truein
{\bf Gordon W. Semenoff}\\
\medskip
{\it Department of Physics, University of British Columbia\\ Vancouver, British
 Columbia, Canada V6T 1Z1}\\

\bigskip
\medskip

{\bf Richard J. Szabo}\\
\medskip
{\it Department of Theoretical Physics, University of Oxford\\ 1 Keble Road,
 Oxford OX1 3NP, U.K.}\\

\vskip 1.5 truein

\end{center}

\begin{abstract}

\baselineskip=12pt

We consider a variation of $O(N)$-symmetric vector models in which the vector
components are Grassmann numbers. We show that these theories generate the same
sort of random polymer models as the $O(N)$ vector models and that they lie in
the same universality class in the large-$N$ limit. We explicitly construct the
double-scaling limit of the theory and show that the genus expansion is an
alternating Borel summable series that otherwise coincides with the topological
expansion of the bosonic models. We also show how the fermionic nature of these
models leads to an explicit solution even at finite-$N$ for the generating
functions of the number of random polymer configurations.

\end{abstract}

\end{titlepage}

\clearpage\newpage

\baselineskip=18pt

The statistical mechanics of randomly branching polymers (sometimes called
discrete filamentary surfaces) is of interest in condensed
matter physics where the connected polymer chains can be
thought of as describing molecules. Random polymers have also been of interest
as examples of random geometry systems which in many cases are exactly
solvable and which have been argued to represent a certain
dimensionally-reduced phase of Polyakov string theory in target space
dimensions $D>1$ \cite{adfj,ddsw}. They therefore serve as toy models for more
complicated higher dimensional problems such as statistical models of
discretized surfaces which have been studied in the context of string
theory and lower dimensional quantum gravity (see \cite{fgz} for a review).
$O(N)$-symmetric vector field theories \cite{hb}
provide non-perturbative models of randomly branching chains where the
order in the $1/N$ expansion coincides with the genus, or number of
loops in the molecules of the ensemble.  These models exhibit phase
transitions in the large-$N$ limit at which infinitely long polymers
dominate the statistical sum and a continuum limit analogous to that
found in the matrix model representation of random surface theories is
reached \cite{adk}--\cite{schel}. The simplicity of the random polymers as
compared to random surfaces allows a more explicit solution, even in dimensions
$D>1$, where some of the ideas about
scaling and other critical behaviour can be tested. These simpler polymer
structures are reflected in the linearity of the vector field theories in
contrast to the non-linearity of matrix ones. The multicritical series
generated by these models give generalized statistical systems which
interpolate between the Cayley tree at one end and the ordinary random walk at
the other \cite{adfj}. Some supersymmetric generalizations of the $O(N)$ vector
model have been studied in \cite{super}.

In this Letter we shall show that a vector theory with purely fermionic degrees
of freedom can also be used to
represent random polymers.  This model exhibits the same critical behaviour in
the large-$N$ limit as the $O(N)$ vector model, but it has a structure which is
in some respects simpler.  In particular, we will
show that its genus expansion is an alternating series which is Borel summable
and that it represents a rare example of a random geometry theory whose
explicit solution can be written down even at finite $N$. It therefore provides
an explicit, well-defined generating function for a given number of polymer
configurations in a random polymer system. It can also be
combined with other polymer models to generate new types of generating
functions for random surface theories. For example, it can be combined with the
$O(N)$ vector model to study random polymer theories
with either only even or only odd genera.

Let us start by reviewing some features of the $O(N)$ vector model with
partition function \cite{nish,divec1,schel}
\beq
Z_S(t,g;N)=\int\prod_{i=1}^Nd\phi_i~\exp\left\{-t\sum_{i=1}^N\phi_i^2+\frac{g}
{N}\left(\sum_{i=1}^N\phi_i^2\right)^2\right\}
\label{bosevec}
\eeq
where the integration is over $\IR^N$. The model (\ref{bosevec}) is invariant
under the orthogonal
transformation $\phi_i\to \sum_{j=1}^NS_{ij}\phi_j$, $S\in O(N)$. This
symmetry restricts the observables of the theory to those which are
functions of $\phi^2\equiv\sum_i\phi_i^2$.  When the coupling
constants $t$ and $g$ are positive, it is straightforward to show that
the formal expression (\ref{bosevec}) counts the number of randomly
branching polymers, both those with a tree-like structure and those
with arbitrarily many loops.  The fact that the integral is divergent
is a reflection of the divergence of the statistical sum.

The statistical sum over polymers coincides with the expansion of the
free energy
\beq
F_S\equiv -\frac{1}{N}\log\left( \frac{ Z_S(t,g;N)}{ Z_S(t,0;N)}\right)
\eeq in Feynman diagrams.  The propagator is
\beq
\left\langle\phi_i\phi_j\right\rangle_S\equiv\frac{\int\prod_kd\phi_k~
\phi_i\phi_j\e^{-t\phi^2}}{\int\prod_kd\phi_k~\e^{-t\phi^2}}=\delta_{ij}~~~~~,
\label{scalarprop}
\eeq
the vertex is \beq \left\langle
\phi_i\phi_j\phi_k\phi_l\right\rangle_S =
\frac{1}{(2t)^2}\left(\delta_{ij}\delta_{kl}+
\delta_{il}\delta_{jk}+\delta_{ik}\delta_{jl}\right) \eeq and the free energy
is the sum of all
connected diagrams with four-point couplings (Figs. 1 and 2). The dual graphs
to these Feynman diagrams, defined by associating a vertex (molecule) in the
center
of each of the scalar loops and lines (bonds) connecting vertices by crossing
each of the Feynman 4-point couplings, are the random walk diagrams shown in
Fig. 2. The number of molecules $n$ appears as the power in
$N^{n-1}$ while the number of bonds $b$ is given by the
power in $(g/Nt^2)^b$ \cite{adk,nish}. This identifies $\frac{1}{N}=\e^\mu$
with the fugacity $\mu$ of the polymer and $\frac{g}{t^2}=\e^{-L}$ with the
length $L$ of the branched chain (the usual action terms for a random
walk model) \cite{adfj}. The vector model partition function is therefore the
generating function for the number of polymer configurations with $b$
bonds and $\ell=b-n+1$ loops. Note that this sum includes the
self-bonding polymers which are generated by Wick contracting several
propagators into single loops (as opposed to multi-loops) and occur in the
expansion only for $\ell\geq1$ (Fig. 2).

\vspace{0.5cm}

\unitlength=1.00mm
\linethickness{0.4pt}
\thicklines
\begin{picture}(130.00,40)
\put(10.00,20.00){\makebox(0,0){$i$}}
\put(10.00,15.00){\line(1,0){20.00}}
\put(30.00,20.00){\makebox(0,0){$j$}}
\put(40.00,15.00){\makebox(0,0){$=~\delta_{ij}$}}
\put(90.00,15.00){\line(-3,-5){5.00}}
\put(90.00,15.00){\line(3,5){5.00}}
\put(90.00,15.00){\line(-3,5){5.00}}
\put(90.00,15.00){\line(3,-5){5.00}}
\put(90.00,15.00){\circle*{2.00}}
\put(84.00,5.00){\makebox(0,0){$l~~$}}
\put(84.00,25.00){\makebox(0,0){$i~~$}}
\put(96.00,5.00){\makebox(0,0){$~~k$}}
\put(96.00,25.00){\makebox(0,0){$~~j$}}
\put(116.00,15.00){\makebox(0,0){$~~~~~~~~~~~=\frac{1}{4t^2}\left(\delta_{ij}
\delta_{kl}+\delta_{il}\delta_{jk}+\delta_{ik}\delta_{jl}\right)$}}
\end{picture}
\begin{description}
\small
\baselineskip=12pt
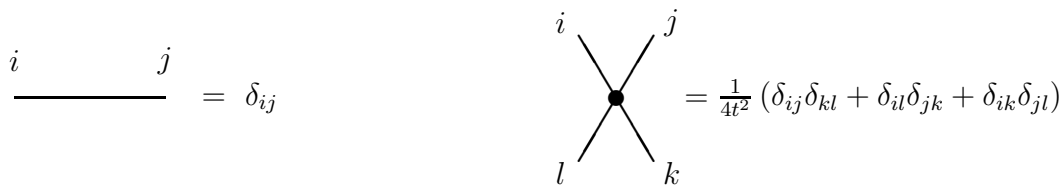
\item[Figure 1:] Feynman rules for the $O(N)$ vector model.
\end{description}

\vspace{0.5cm}

\unitlength=1.00mm
\linethickness{0.4pt}
\thicklines
\begin{picture}(140.00,60.00)
\put(0.00, 35.00){\makebox(0,0)[l]{$F_S=$}}
\put(16.0,35.00){\circle{7.00}}
\put(19.50,35.00){\circle*{2.00}}
\put(23.00,35.00){\circle{7.00}}
\put(26.50,35.00){\circle*{2.00}}
\put(30.00,35.00){\circle{7.00}}
\thinlines
\put(16.00,35.00){\circle*{1.00}}
\put(16.00,35.00){\line(1,0){7.00}}
\put(23.00,35.00){\circle*{1.00}}
\put(23.00,35.00){\line(1,0){7.00}}
\put(30.00,35.00){\circle*{1.00}}
\put(33.00,35.00){\makebox(0,0)[l]{$~~~+~~~$}}
\thicklines
\put(48.0,35.00){\circle{7.00}}
\put(51.50,35.00){\circle*{2.00}}
\put(55.00,35.00){\circle{7.00}}
\put(58.50,35.00){\circle*{2.00}}
\put(62.00,35.00){\circle{7.00}}
\put(65.50,35.00){\circle*{2.00}}
\put(69.00,35.00){\circle{7.00}}
\put(72.50,35.00){\circle*{2.00}}
\put(76.00,35.00){\circle{7.00}}
\thinlines
\put(48.00,35.00){\circle*{1.00}}
\put(48.00,35.00){\line(1,0){7.00}}
\put(55.00,35.00){\circle*{1.00}}
\put(55.00,35.00){\line(1,0){7.00}}
\put(62.00,35.00){\circle*{1.00}}
\put(62.00,35.00){\line(1,0){7.00}}
\put(69.00,35.00){\circle*{1.00}}
\put(69.00,35.00){\line(1,0){7.00}}
\put(76.00,35.00){\circle*{1.00}}
\put(79.00,35.00){\makebox(0,0)[l]{$~~~+~~~$}}
\thicklines
\put(55.00,38.50){\circle*{2.00}}
\put(55.00,42.00){\circle{7.00}}
\put(55.00,45.50){\circle*{2.00}}
\put(55.00,49.00){\circle{7.00}}
\thinlines
\put(55.00,35.00){\line(0,1){7.00}}
\put(55.00,42.00){\circle*{1.00}}
\put(55.00,42.00){\line(0,1){7.00}}
\put(55.00,49.00){\circle*{1.00}}
\thicklines
\put(69.00,31.50){\circle*{2.00}}
\put(69.00,28.00){\circle{7.00}}
\thinlines
\put(69.00,35.00){\line(0,-1){7.00}}
\put(69.00,28.00){\circle*{1.00}}
\thicklines
\put(94.0,35.00){\circle{7.00}}
\put(97.50,35.00){\circle*{2.00}}
\put(101.00,35.00){\circle{7.00}}
\put(104.50,35.00){\circle*{2.00}}
\put(108.00,35.00){\circle{7.00}}
\put(111.50,35.00){\circle*{2.00}}
\put(115.00,35.00){\circle{7.00}}
\put(118.50,35.00){\circle*{2.00}}
\put(122.00,35.00){\circle{7.00}}
\put(125.50,35.00){\circle*{2.00}}
\put(129.00,35.00){\circle{7.00}}
\thinlines
\put(94.00,35.00){\circle*{1.00}}
\put(94.00,35.00){\line(1,0){7.00}}
\put(101.00,35.00){\circle*{1.00}}
\put(101.00,35.00){\line(1,0){7.00}}
\put(108.00,35.00){\circle*{1.00}}
\put(108.00,35.00){\line(1,0){7.00}}
\put(115.00,35.00){\circle*{1.00}}
\put(115.00,35.00){\line(1,0){7.00}}
\put(122.00,35.00){\circle*{1.00}}
\put(132.00,35.00){\makebox(0,0)[l]{$~~~+~~\dots$}}
\put(122.00,35.00){\line(1,0){7.00}}
\put(129.00,35.00){\circle*{1.00}}
\thicklines
\put(101.00,42.00){\circle{7.00}}
\put(101.00,38.50){\circle*{2.00}}
\put(101.00,49.00){\circle{7.00}}
\put(101.00,45.50){\circle*{2.00}}
\thinlines
\put(101.00,35.00){\line(0,1){7.00}}
\put(101.00,42.00){\circle*{1.00}}
\put(101.00,42.00){\line(0,1){7.00}}
\put(101.00,49.00){\circle*{1.00}}
\thicklines
\put(104.50,49.00){\circle*{2.00}}
\put(108.00,49.00){\circle{7.00}}
\put(111.50,49.00){\circle*{2.00}}
\put(115.00,49.00){\circle{7.00}}
\put(118.50,49.00){\circle*{2.00}}
\put(122.00,49.00){\circle{7.00}}
\thinlines
\put(101.00,49.00){\line(1,0){7.00}}
\put(108.00,49.00){\circle*{1.00}}
\put(108.00,49.00){\line(1,0){7.00}}
\put(115.00,49.00){\circle*{1.00}}
\put(115.00,49.00){\line(1,0){7.00}}
\put(122.00,49.00){\circle*{1.00}}
\put(122.00,49.00){\line(0,-1){7.00}}
\put(122.00,42.00){\circle*{1.00}}
\put(122.00,42.00){\line(0,-1){7.00}}
\thicklines
\put(122.00,42.00){\circle{7.00}}
\put(122.00,45.50){\circle*{2.00}}
\put(122.00,38.50){\circle*{2.00}}
\thicklines
\put(10.00,5.00){\makebox(0,0)[l]{$+$}}
\put(19.00,5.00){\circle{7.00}}
\put(19.00,5.00){\circle*{1.00}}
\put(22.00,5.00){\makebox(0,0)[l]{$~~~+$}}
\put(41.00,8.00){\circle{20.00}}
\put(41.00,14.50){\circle*{2.00}}
\put(41.00,12.00){\circle{5.00}}
\put(51.00,5.00){\makebox(0,0)[l]{$~~~+$}}
\thinlines
\put(41.00,9.50){\circle{10.00}}
\put(41.00,4.50){\circle*{1.00}}
\thicklines
\put(71.00,8.00){\circle{20.00}}
\put(71.00,14.50){\circle*{2.00}}
\put(71.00,13.00){\circle{3.00}}
\put(71.00,11.50){\circle*{2.00}}
\put(71.00,10.00){\circle{3.00}}
\thinlines
\put(71.00,9.00){\circle{11.00}}
\put(71.00,7.50){\circle{8.00}}
\put(71.00,3.50){\circle*{1.00}}
\put(81.00,5.00){\makebox(0,0)[l]{$~~~+~~\dots$}}
\end{picture}
\begin{description}
\small
\baselineskip=12pt
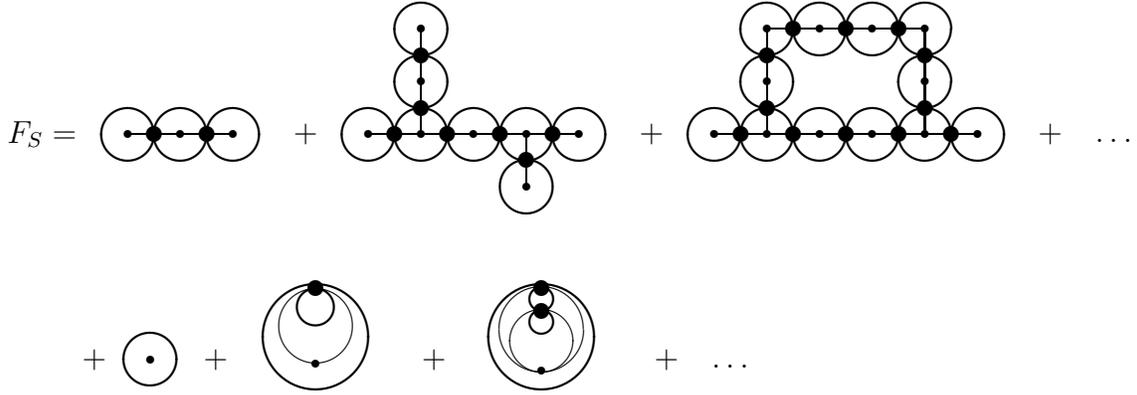
\item[Figure 2:] Feynman diagram expansion of the $D=0$ quantum field theory
(\ref{bosevec}). The thick lines and vertices represent the Feynman graphs,
while the thin lines and vertices represent the dual graphs which form polymer
networks (or random walks). The first set of diagrams represent those types of
branching polymers associated with the higher orders of $N$ in vertex
contractions such as $\langle(\phi^2)^2\rangle_S=(N^2+2N)/4t^2$ in Fig. 1 (so
that $n>1$ and bonds connect different molecules), while the second set of
diagrams representing the self-bonding polymers are associated with the lower
powers of $N$ (so that $\ell=b$).
\end{description}

\vspace{0.5cm}

{}From an analytic point of view, the perturbative expansion of the
partition function
\beq \frac{Z_S(t,g;N)}{Z_S(t,0;N)}=
\sum_{n\geq0}\frac{1}{n!}\left(\frac{g}{N}\right)
^n\left\langle\left(\phi^2\right)^{2n}\right\rangle_S \eeq is completely
determined by the Gaussian moments \beq
\left\langle\left(\phi^2\right)^{2n}\right\rangle_S
\equiv\frac{\int\prod_id\phi_i~
(\phi^2)^{2n}\e^{-t\phi^2}}{\int\prod_id\phi_i
{}~\e^{-t\phi^2}}=t^{N/2}\left(\frac{ \partial^2}{\partial
t^2}\right)^nt^{-N/2}
\label{gaussmom}\eeq as \beq
\frac{Z_S(t,g;N)}{Z_S(t,0;N)}=\sum_{n=0}^{N_\Lambda}
\frac{(N+4n-2)!!}{2^{2n} n!(N-2)!!}\left(\frac{g}{Nt^2}\right)^n
\label{zscutoff}\eeq
Here we have introduced an ``ultraviolet" cutoff $N_\Lambda\in\IZ^+$
to make the partition function well-defined. In the limit
$N_\Lambda\to\infty$, the series (\ref{zscutoff}) is a non-Borel
summable asymptotic series reflecting the divergence of the original
integral and also the divergence of the statistical sum.
As is familiar from the study of random surfaces \cite{fgz}, even though the
series is divergent, if arranged as a power series in $1/N$, rather
than $g$, the terms in this series are individually convergent and it
is the sum over genera $\ell$ which is asymptotic
\cite{adfj},\cite{adk}--\cite{schel}. With the cutoff $N_\Lambda$ in
(\ref{zscutoff}), the partition function is an analytic function of $N$, but it
is only well-defined at $N=\infty$ when this ultraviolet cutoff is removed.

The large $N$ expansion is a saddle point computation of the integral
(\ref{bosevec}). After integration over angular variables,
the partition function (\ref{bosevec}) can be written in terms of the
radial coordinate in Euclidean $N$-space as \beq
Z_S(t,g;N)=\frac{2\pi^{N/2}}{\Gamma(N/2)}\int_0^\infty
d\phi~\phi^{N-1}\e^{-t\phi^2+\frac{g}{N}\phi^4}
\label{zsradial}\eeq
In the infinite-$N$ limit, which counts the tree-graphs ($\ell=0$),
the integral (\ref{zsradial}) can be evaluated using the saddle-point
approximation. Rescaling $\phi\to\phi/\sqrt{N}$ the stationary condition for
the effective action $-Nt\phi^2+gN\phi^4+N\log\phi$ in (\ref{zsradial})
is \beq 2t\phi^2-4g\phi^4=1
\label{statcond}\eeq
The solution of (\ref{statcond}) which is regular at $g=0$ and which minimizes
the effective action is
\beq
\phi_0^2=\frac{t}{4g}\left(1-\sqrt{1-\frac{4g}{t^2}}\right)
\label{saddlepoint}\eeq
The tree-level free energy is then the value of the effective action
in (\ref{zsradial}) evaluated at the saddle-point (\ref{saddlepoint}),
\beq F_S^{(0)}\equiv\lim_{N\to\infty}-\frac{1}{N}\log Z_S(t,g;N)=\frac{1}{2}
\left\{\frac{1}{2}+\frac{t}{4g}\left(t-\sqrt{t^2-4g}\right)-\log
\left(\frac{t}{4g}-\frac{1}{4g}\sqrt{t^2-4g}\right)\right\}
\label{fstree}\eeq
This free energy becomes non-analytic at the critical point
$g=g_c\equiv t^2/4$ where the 2 solutions of the quadratic equation
(\ref{statcond}) coalesce. There the minimum of the effective action in
(\ref{zsradial}) disappears and it becomes unbounded \cite{divec1,zinn1}, so
that the saddle point
solution is no longer valid. There is a second order phase transition at
the coupling $g=g_c$ with susceptibility exponent
$\gamma^{(0)}=\frac{1}{2}$ \cite{adfj,adk,nish}. This critical point is
identified as the
``continuum" limit of the random polymer theory where the number of
branches, and hence the lengths, of the tree-graphs becomes
infinite. The higher-loop contributions (molecular networks) can be
found in \cite{schel} and their ``continuum" limit is associated with
an infinite number of molecules, thus tracing out a continuum filamentary
surface \cite{adfj,adk}.

Random polymer models can also be generated by the fermonic vector
model with partition function \beq Z_F(t,g;N)=\int
d\psi~d\bar\psi~\e^{t\ps2-\frac{g}{2N}(\ps2)^2}
\label{part0}\eeq
where $\psi_i,\bar\psi_i$, $i=1,\dots,N$, are independent anticommuting
nilpotent variables, $\ps2\equiv\sum_{i=1}^N\bar{\psi}_i\psi_i$ and the
integration measure
$d\psi~d\bar\psi\equiv\prod_{i=1}^Nd\psi_i~d\bar\psi_i$ is defined
using the usual Berezin rules for integrating Grassmann variables,
$\int d\psi_i~\psi_i=1$, $\int d\psi_i~1=0$. The model (\ref{part0}) possesses
a continuous symmetry, $\psi_i\to
\sum_jU_{ij}\psi_j,\bar\psi_i\to\sum_j\bar\psi_j(U^{-1})_{ji}$ with $U\in
GL(N,\IC)$. There is a further discrete symmetry under the ``chiral"
transformation
$\psi_i\rightarrow\bar\psi_i,\bar\psi_i\rightarrow -\psi_i$ for any
$i$, which is the analog of the reflection symmetry $\phi_i\to-\phi_i$ of the
$O(N)$ vector model. Together, these two symmetries restrict the observables of
the model to those which are functions only of $\ps2$.  We shall now show that
the model (\ref{part0}) possesses a random geometry interpretation and
critical behaviour similar to that of the $O(N)$ vector model. A main
difference with the $O(N)$ vector model is that the integration over
Grassmann variables in the generating function (\ref{part0}) for the
polymers is a well-defined finite polynomial in the coupling constants
$g$ and $t$ because nilpotency of the components $\psi_i$ and $\bar\psi_i$
implies
that $(\ps2)^{N+1}=0$. The dimension $N$ itself provides a cutoff on the number
of terms (and polymers) in the Feynman diagram expansion of (\ref{part0}).
Here, rather than making the partition function integration well-defined as in
the bosonic case, the large-$N$ limit is needed to
generate the full ensemble of randomly-branched chains.

The Feynman diagrams for the fermion vector theory (\ref{part0}) have
propagator
\beq \left\langle\bar\psi_i\psi_j\right\rangle_F\equiv\frac{\int
d\psi~d\bar\psi~\bar\psi_i\psi_j\e^{t\ps2}}{\int
d\psi~d\bar\psi~\e^{t\ps2}}=\delta_{ij} \eeq
and the four-Fermi interaction vertex is
\beq
\left\langle\bar\psi_i\psi_j\bar\psi_k\psi_l\right\rangle_F
=\frac{1}{t^2}\left(\delta_{ij}\delta_{kl}-\delta_{il}\delta_{jk}\right)
\label{4fvertex}\eeq
These Feynman rules have the same graphical representation shown in Fig. 1 with
a left-handed orientation for the lines (ingoing lines into a vertex for
$\bar\psi_i$ components and outgoing lines for $\psi_j$ components). Now the
graphs are formed from all connected 4-point diagrams which preserve this
orientation. The Feynman rules also
associate a factor of $-1$ to each fermion loop in a Feynman graph.

The perturbative expansion of (\ref{part0}) \beq
\frac{Z_F(t,g;N)}{Z_F(t,0;N)}=\sum_{n\geq0}\frac{(-1)^n}{n!}\left(\frac{g}{2N}
\right)^n\left\langle(\ps2)^{2n}\right\rangle_F \eeq is completely
determined by the normalized Gaussian moments \beq
\left\langle(\ps2)^{2k}\right\rangle_F\equiv\frac{\int
d\psi~d\bar\psi~(\ps2)^{2k}\e^{t\ps2}}{\int
d\psi~d\bar\psi~\e^{t\ps2}}=t^{-N}\left(\frac{\partial^2}{\partial
t^2}\right)^kt^N
\label{gaussvecmom}\eeq
to be \beq \frac{Z_F(t,g;N)}{Z_F(t,0;N)}=\sum_{n=0}^{N_2}(-1)^n
\frac{N!}{n!(N-2n)!}\left(\frac {g}{2Nt^2}\right)^n
\label{fermpert}\eeq
where $N_2=N/2$ (respectively $(N-1)/2$) when $N$ is even (odd). The
perturbation series is a finite sum which represents the same sort of random
walk distribution as shown in Fig. 2 except that it only includes polymers with
up to $N_2$ bonds. From a diagrammatic point of
view, the alternating nature of the series arises from the minus sign
associated with fermion loops.  Term by term, this series can be made
identical with the terms of same order in $g$ in (\ref{zscutoff}) by
the analytical continuation $N\rightarrow -N/2$ in (\ref{fermpert}) (so that
$N_\Lambda=N_2$ in (\ref{zscutoff})). The factor of 2 is associated with the
doubling of degrees of freedom in the fermionic case. Thus, after the
substitution
$N\rightarrow N/2$, the large $N$ expansion of the fermionic vector
model is identical to that of the $O(N)$ vector model {\it except} that
it is an alternating series in $1/N$.  The coefficient of $1/N^k$ in
the former and $1/(-N)^k$ in the latter are identical. Now, however, the
alternating nature of the fermionic vector series makes its $N\to\infty$ limit
Borel summable, and as such it defines a better behaved statistical theory.

To explicitly carry out the $\frac{1}{N}$-expansion of the fermionic vector
model, we introduce a scalar Hubbard-Stratonovich field $\varphi$ by inserting
the identity $1=\int d\varphi~\e^{-\frac{g}{2N}(\varphi+i\ps2)^2}$
into the partition function integral (\ref{part0}) to write it as
\beq
Z_F(t,g;N)=\int d\varphi~\e^{-g\varphi^2/2N}\int
d\psi~d\bar\psi~\e^{(t-ig\varphi/N)\ps2}=N!\int
d\varphi~(t-ig\varphi/N)^N\e^{-g
\varphi^2/2N}
\label{zfhub}\eeq
When $N\to\infty$ the integral in (\ref{zfhub}) is determined by the
saddle-point value of $\varphi$. Rescaling $\varphi\to\varphi/N$, this can be
found from the stationary condition for the effective action
\beq
S(\varphi)\equiv-Ng\varphi^2/2+N\log(t-ig\varphi)
\label{effaction}\eeq
appearing in (\ref{zfhub}) which is \beq tg\varphi-ig^2\varphi^2+ig=0
\label{fermsaddle}\eeq
The solution of (\ref{fermsaddle}) which is regular at $g=0$ is \beq
\varphi_0=\frac{t}{2ig}\left(1-\sqrt{1-\frac{4g}{t^2}}\right)
\label{hub0}\eeq
Substituting (\ref{hub0}) into (\ref{effaction}) we get the tree-level
fermionic free energy
\beq
F_F^{(0)}=\frac{1}{2}-\frac{t}{4g}\left(t-\sqrt{t^2-4g}\right)-\log\left
(\frac{t}{2}+\sqrt{t^2-4g}\right)
\label{fermfree0}\eeq
Note that the saddle points in (\ref{saddlepoint}) and (\ref{hub0}) are related
by the correspondence $\phi^2=\frac{i}{2}\varphi$ which makes more precise the
analytical continuation discussed above. Notice also that the free energy
(\ref{fermfree0}) is related to the free energy (\ref{fstree}) of the scalar
model as $F_F^{(0)}=1-2F_S^{(0)}$, as anticipated since the
large-$N$ limit of the 2 models represents the same combinatorical problem
of enumerating tree-graphs.

The higher-loop contributions (which count the polymer networks with a given
number of molecules) can be found by carrying out the saddle point calculation
of the integral (\ref{zfhub}) to higher orders. For this, we decompose the
Hubbard-Stratonovich field as $\varphi=\varphi_0+\varphi_q$ and expand the
action (\ref{effaction}) in a Taylor series about the saddle-point value
(\ref{hub0}) in terms of the fluctuation fields $\varphi_q$. Using the
saddle-point equation (\ref{fermsaddle}) when evaluating the higher-order
derivatives $S^{(n)}(\varphi_0)$, this Taylor series is found to be
\beq
S(\varphi)=S(\varphi_0)-\frac{N}{2}\left(g+g^2\varphi_0^2\right)\varphi_q^2-N
\sum_{n=3}^\infty\frac{(-g\varphi_0)^n}{n}\varphi_q^n
\label{actiontaylor}\eeq
The genus 1 free energy is then obtained from the fluctuation determinant that
arises from Gaussian integration over the quadratic part in $\varphi_q$ of
(\ref{actiontaylor}),
\beq
F_F^{(1)}=\frac{1}{2N}\log\left(g+g^2\varphi_0^2\right)=\frac{1}{2N}\log\left(
2g-\frac{t^2}{2}+\frac{t}{2}\sqrt{t^2-4g}\right)
\label{fermfree1}\eeq
which also agrees with the 1-loop free energy of the $O(N)$ vector model
\cite{schel}.

The $\frac{1}{N}$-expansion of the fermionic free energy also becomes
non-analytic at the critical point $g=g_c=t^2/4$. It exhibits the same critical
behaviour as the $\phi^4$ theory above and it therefore lies in the same
universality class as this statistical model. It is straightforward to carry
out the double-scaling limit of the fermionic vector model in much the same way
as in the bosonic case \cite{nish,divec1,schel}. This limit is associated with
the continuum limit of
the polymer network at $N\to\infty$, $g\to g_c$ in such a way that a coherent
contribution from all orders of the perturbative and $1/N$ expansions is
obtained. We approach the critical point $g_c$ by defining a dimensionless
``lattice spacing'' $a$ by $at^2=g_c-g$ and taking the continuum limit $a\to0$.
 With the rescalings mentioned above, the contribution to an arbitrary
$\ell$-loop vacuum diagram with $\ell\geq1$ is
$N^{-b}(\sqrt{a})^{-b}N^n=N^{1-\ell}(\sqrt{a})^{1-\ell-n}$. As shown in
\cite{divec1}, the maximum number of vertices that a 4-point polymer diagram
with $\ell$ loops can have is $n=2(\ell-1)$, so that the most singular
behaviour of an $\ell$-loop diagram in the continuum limit $a\to0$ is
$(Na^{3/2})^{1-\ell}$. The proper continuum limit wherein a finite contribution
from arbitrary genus polymers is obtained is thus the ``double-scaling" limit
where $N\to\infty$ and $a\to0$ in a correlated fashion so that the renormalized
``cosmological constant" (or ``linear string tension") $\Lambda\equiv Na^{3/2}$
remains finite. The double scaling limit enables an explicit construction of
the genus expansion of the continuum polymer theory from the vector model.

We can now take the double scaling limit of the partition function
(\ref{part0}) and write it as a loop expansion in the linear string tension
$\Lambda$. However, as noted for the $O(N)$ vector model \cite{divec1}, the
tree and 1-loop contributions are singular in this limit. The saddle-point
value (\ref{hub0}) can be written in terms of the lattice spacing as
$g\varphi_0=(2/it)(1-2\sqrt{a})$, from which it follows that the genus 0 and 1
free energies (\ref{fermfree0}) and (\ref{fermfree1}) are given by
\beq\new{\begin{array}{c}
NF_F^{(0)}\sim-\frac{N}{2}-N\log\left(\frac{t}{2}\right)+6N^{1/3}\Lambda^{2/3}-
\frac{32}{3}\Lambda~~~,~~~
NF_F^{(1)}\sim\frac{1}{2}\log\left(\frac{t^2}{2}\right)+\frac{1}{2}\log
\left(\frac{2\Lambda^{1/3}}{N^{1/3}}\right)
\end{array}}
\label{nonuniv}\eeq
in the continuum limit $a\to0$. The $\Lambda$-dependent terms in
(\ref{nonuniv}) diverge in the double scaling limit and represent a
non-universal behaviour of the statistical polymer system. The tree-level and
one-loop order Feynman diagrams should therefore be subtracted in the
definition of the double-scaling limit leading to a renormalized partition
function $Z_R(\Lambda,t)$ that only contains contributions from the $\ell$-loop
diagrams with $\ell\geq2$.

This renormalized partition function is obtained by integrating over that part
of the action involving $n\geq3$ vertices in the fluctuation field $\varphi_q$
weighted against the Gaussian form in (\ref{actiontaylor}). To pick out the
finite contribution in the double-scaling limit, we rescale the fluctuation
field as $\varphi_q\to2N^{1/3}\Lambda^{1/6}g_c^{-1/2}\varphi_q$ and note that
with this rescaling we have
\beq
-\frac{N}{2}\left(g+g^2\varphi_0^2\right)\varphi_q^2\to-\frac{1}{2}\varphi_q^2
{}~~~,~~~-N(-g\varphi_0)^n\varphi_q^n\to-\frac{N(g_c)^{n/2}}{2^n}
\left(\frac{it}{2}\right)^nN^{-n/3}\Lambda^{-n/6}\varphi_q^n
\label{rescaling}\eeq
in the continuum limit $a\to0$. The $n\geq4$ vertex terms in
(\ref{rescaling}) vanish in the double scaling limit, and therefore the exact
renormalized partition function in the double scaling limit is (up to
irrelevant normalization factors)
\beq
Z_R(\Lambda,t)=\frac{\int
d\varphi_q~\e^{-\frac{1}{2}\varphi_q^2+\frac{it^6}{512}\Lambda^{-1/2}\varphi_q
^3}}{\int
d\varphi_q~\e^{-\frac{1}{2}\varphi_q^2}}=\sum_{k=0}^\infty\frac{i^kt^{6k}}
{k!(512)^k}~\Lambda^{-k/2}\left(\left\langle\phi^{3k}\right\rangle_S
\biggm|_{N=1,t=\frac{1}{2}}\right)
\label{zren}\eeq
The Gaussian moments in (\ref{zren}) can be evaluated as in (\ref{gaussmom}).
The odd moments vanish, while the even moments yield a factor $(3k-1)!!$. Thus
the double-scaled renormalized partition function admits the exact genus
expansion
\beq
Z_R(\Lambda,t)=\sum_{\ell=0}^\infty(-1)^\ell
{}~t^{12\ell}~\frac{(6\ell-1)!!}{(2\ell)!(512)^{2\ell}}~\Lambda^{-\ell}
\label{renloopexp}\eeq

The partition function (\ref{renloopexp}) has a similar structure as that in
the $O(N)$ vector model where the genus expansion is an asymptotic series with
zero radius of convergence \cite{divec1}. In the fermionic case, however, the
genus expansion
is an alternating sum, and is therefore Borel summable. The convergence of the
sum over genera is easily seen in the integral expression (\ref{zren}) where
the unbounded cubic term contains a factor of $i$ which makes the overall
integration there finite. The Borel summability of the genus expansion is a
feature unique to the fermionic models that does not usually occur for
random geometry theories. In this sense, the fermionic vector model represents
some novel discretized surface theory in which the topological expansion
uniquely specifies the generating function of the statistical theory.

The two functions
\beq
Z_\pm(t,g;N)=\sqrt{Z_S(t,g;N)Z_F(t,g;N/2)^{\pm1}}
\eeq
are partition functions of random polymer models where, in the case of the
+ sign the genera are restricted to be even and in the case of the --
sign the genera are odd. The resulting statistical theory is not Borel
summable, but it does represent a new sort of generating function for
``reduced" polymer systems. In these reduced statistical models, at least part
of the non-universal genus 0 or 1 double-scaling behaviour is removed. The
convergence properties of the fermionic vector
models can therefore be exploited to combine them with other vector models and
generate unusual statistical models of random polymer systems. It would be
interesting to give the fermionic nature of these models an interpretation in
terms of random surfaces directly. The identification $-\frac{1}{N}=\e^\mu$ in
the fermionic case suggests that the associated random polymer theory has a
complex-valued ``fugacity" $\mu=i\pi+\mu_0$, $\mu_0\in\IR$, with
doubly-degenerate degrees of freedom at each vertex. Heuristically, the complex
fugacity $\mu$ can be thought of as a rotation of the real-valued one $\mu_0$
required to compensate the extra vertex degree of freedom that does not appear
in the bosonic models. From an analytic point of view, the genus sum alternates
relative to that of the $O(N)$ vector model because the saddle-point
(\ref{hub0}) is imaginary in the fermionic vector model (\ref{part0}) so that
its saddle-point expansion is the analytical continuation
$\phi_0^2=\frac{i}{2}\varphi_0$ of that for the scalar model.

The above analysis can be straightforwardly generalized to an interaction of
the form $g(\ps2)^K$, which will then represent a random polymer model with up
to $2K$-valence vertices. The critical behaviour is the same as that in a
$\phi^{2K}$ scalar vector model and leads to the same susceptibility exponent
$\gamma^{(0)}=\frac{1}{2}$, i.e. such a theory of random polymers is universal.
To generate more complicated polymer models, for instance those with matter
degrees of freedom at the vertices of the discretization \cite{adfj}, one must
study vector
models with more complicated interactions. One particularly interesting aspect
of the fermionic vector model is the extent to which these more complex models
can be solved explicitly. Consider the more general fermionic vector model with
partition function
\beq
Z_0=\int d\psi~d\bar\psi~\e^{NV(\ps2)}
\label{partgen}\eeq
where $V(z)$ is some ``potential" function. The integration in (\ref{partgen})
is well-defined and finite if $\e^{NV(z)}$ has a well-defined Taylor expansion
to order $N$ in the variable $z$. In that case, the partition function
(\ref{partgen}) at finite $N$ can be formally evaluated
by inserting the delta function $1=\int dz~\delta(z-\bar\psi\psi)$ and using
the identity \beq
\int\frac{dw}{2\pi}~ \int d\psi~d\bar\psi~\e^{iw(z-\ps2)}=
\int\frac{dw}{2\pi}~N!(-iw)^N
\e^{iwz}=N!\left(-\frac{\partial} {\partial z}\right)^N\delta(z)
\label{partid}\eeq
to obtain \beq Z_0=N!\cdot \left(\frac{\partial}{\partial
z}\right)^N\e^{NV(z)}\biggm\vert_{z=0}
\label{z0}\eeq
Similarly, the correlators for $N$ finite are given by \beq
\left\langle(\ps2)^n\right\rangle\equiv\frac{\int
d\psi~d\bar\psi~(\ps2)^n\e^{NV(\ps2)}}{\int
d\psi~d\bar\psi~\e^{NV(\ps2)}}=\frac{\left(\frac{\partial}{\partial
z}\right)^Nz^n \e^{NV(z)}}{\left(\frac{\partial}{\partial
z}\right)^N\e^{NV(z)}}\biggm|_{z=0} \eeq
The generating function and observables for these fermionic polymer models can
therefore always be explicitly and uniquely specified. It is straightforward to
check that (\ref{z0}) reduces to (\ref{fermpert}) when the potential is the
four-Fermi interaction that we studied above.

To treat the model (\ref{partgen}) at $N=\infty$, we use the first part of the
identity (\ref{partid}) to write the partition function as
\beq
Z_0=\frac{(-i)^NN!}{2\pi}\int dz~dw~\e^{NV(z)+iwz+N\log w}
\label{partlargeN}\eeq
If the potential is a polynomial of degree $m$,
\beq
V(z)=\sum_{k=1}^mg_kz^k
\label{genpot}\eeq
then we can rescale $z\to z/N$ and the coupling constants $g_k\to N^k\cdot g_k$
simultaneously so that the effective action in (\ref{partlargeN}) is
$NV(z)+iNwz+N\log w$. The integral (\ref{partlargeN}) at large-$N$ is
determined by the saddle-point value of this effective action. In the
2-dimensional complex space of the variables $w$ and $z$, the stationary
conditions are
\beq
V'(z)+iw=0~~~~~,~~~~~iz+1/w=0
\label{2saddles}\eeq
which can be combined into the single equation
\beq
zV'(z)=1
\label{ONstatgen}\eeq

The equation (\ref{ONstatgen}) is identical to the stationary condition for the
$O(N)$ vector model defined with potential $V(\phi^2)$ \cite{divec1,zinn1}.
Thus the critical behaviour of the fermionic vector model (\ref{partgen}) is
the same as that for the $O(N)$ vector model with the same polynomial potential
(\ref{genpot}). The genus expansion is generated by the 2-dimensional
saddle-point evaluation of the integral (\ref{partlargeN}). The imaginary
saddle-point values given by (\ref{2saddles}) will lead to an alternating genus
expansion in the double-scaling limit for the fermionic theory, leading to a
Borel summable polymer model, in contrast to the scalar case. In this case the
critical point is again that point in coupling constant space where the
function $zV'(z)$ vanishes. For a potential of the form (\ref{genpot}), we can
adjust the coupling constants in such a way that the critical point is a zero
of $zV'(z)$ of order $m$. The leading singularity of the free energy will then
be $a^{(m+1)/m}$ \cite{nish,divec1} which leads to the critical susceptibility
exponent $\gamma^{(0)}=1-1/m$, $m=2,3,\dots$. This is the multicritical series
for generalized random polymer systems in dimension $D\geq0$ \cite{adfj} which
interpolates between the Cayley tree at $m=\infty$ with $\gamma^{(0)}=1$ and
the ordinary random walk we discussed earlier at $m=2$ with
$\gamma^{(0)}=\frac{1}{2}$ (the $N^0$-component of the vector model free energy
represents the self-avoiding random walk). In the $O(N)$ models, the former
case would represent a phase of bosonic strings in target space dimension
$D\geq1$ while the latter case would represent a phase of pure 2-dimensional
quantum gravity. In the general case, the potential (\ref{genpot}) leads to
discrete filamentary surfaces which have vertices of even valence up to $2m$.
It would be interesting to determine what physical systems the fermionic vector
models represent in the continuum limit.

We see therefore that the fermionic vector model partition function is a
well-defined, finite generating function for polymer configurations which can
always (for any $N$) be evaluated exactly. The nature of its genus expansion,
i.e. that it has an alternating, Borel summable double scaling series, agrees
with conjectures concerning the nature of the topological expansion in
fermionic matrix models which have been studied in \cite{mz} as alternative
random surface theories to the conventional Hermitian matrix models. There it
is conjectured that the genus expansion is an alternating series which may be
Borel summable but otherwise coincides with the usual Painlev\'e expansion
\cite{fgz}. This is expected to be only true for odd polynomial potentials as
it is only in that case that the matrix model possesses a chiral symmetry and
imaginary endpoints for the support of the spectral distribution. In the case
of the simpler fermionic vector models the partition function is always
invariant under chiral transformation of the fermionic vector components.
Furthermore, in the case of fermionic matrix models the correspondence with a
scalar theory is more complicated -- a polynomial fermion model can be
analytically continued to a Hermitian matrix model with a generalized Penner
potential which consists of a logarithmic plus polynomial potential \cite{mz}.
It should therefore represent a Borel summable generating function for the
virtual Euler characteristics of the discretized moduli spaces of Riemann
surfaces (rather than just the generating function for a random surface
triangulation itself). In the vector case, the fermionic model represents the
same type of random surface theory as the corresponding $O(N)$ vector field
theory. Nevertheless, further study of the more technical features of fermionic
vector models, such as an analysis of the Schwinger-Dyson (loop) equations and
the connection with integrable hierarchies, may help in understanding what the
corresponding structures will look like in the case of fermionic matrix models.
In the vector case, they will generate examples of fermionic models which are
explicitly solvable.

\bigskip

We wish to thank E. Elizalde for comments on the manuscript, and S. de Queiroz
and J. Wheater for interesting discussions.


\newpage

\begin{thebibliography}{10}

\baselineskip=12pt

\bibitem{adfj} J. Ambj\o rn, B. Durhuus, J. Fr\"ohlich and P. Orland, Nucl.
Phys. {\bf B270} [FS16] (1986), 457; J. Ambj\o rn, B. Durhuus and J.
Fr\"ohlich, Nucl. Phys. {\bf B275} [FS17] (1986), 161; J. Ambj\o rn, B. Durhuus
and T. J\'onsson, Phys. Lett. {\bf B244} (1990), 403; S. Nishigaki, Mod. Phys.
Lett. {\bf A9} (1994), 631

\bibitem{ddsw} S. R. Das, A. Dhar, A. M. Sengupta and S. R. Wadia, Mod. Phys.
Lett. {\bf A5} (1990), 1041; Yu. Makeenko, Intern. J. Mod. Phys. {\bf A10}
(1995), 2615; in Proc. 29th Intern. Ahrenshoop Symp. (1995), to appear

\bibitem{fgz} P. Di Francesco, P. Ginsparg and J. Zinn-Justin, Phys. Rep. {\bf
254} (1995), 1

\bibitem{hb} L. Dolan and R. Jackiw, Phys. Rev. {\bf D9} (1974), 3320; S.
Coleman, R. Jackiw and H. D. Politzer, Phys. Rev. {\bf D10} (1974), 2491; R. G.
Root, Phys. Rev. {\bf D10} (1974), 3322; M. Kobayashi and T. Kugo, Prog. Theor.
Phys. {\bf 54} (1975), 1537; L. F. Abbott, J. S. Kang and H. J. Schnitzer,
Phys. Rev. {\bf D13} (1976), 2212; S. Hikami and E. Br\'ezin, J. Phys. {\bf
A12} (1979), 759

\bibitem{adk} B. Duplantier and I. K. Kostov, Nucl. Phys. {\bf B340} (1990),
491; A. Anderson, R. C. Myers and V. Periwal, Nucl. Phys. {\bf B360} (1991),
463

\bibitem{nish} S. Nishigaki and T. Yoneya, Nucl. Phys. {\bf B348} (1991), 787

\bibitem{divec1} P. Di Vecchia, M. Kato and N. Ohta, Nucl. Phys. {\bf B357}
(1991), 495

\bibitem{zinn1} J. Zinn-Justin, Phys. Lett. {\bf B257} (1991), 335; P. Di
Vecchia and M. Moshe, Phys. Lett. {\bf B300} (1993), 49; G. Eyal, M. Moshe, S.
Nishigaki and J. Zinn-Justin: The $O(N)$ Vector Model in the Large $N$ Limit
Revisited, Saclay preprint SPhT/95-132 (1996)

\bibitem{schel} S. Schelstraete and H. Verschelde, Phys. Lett. {\bf B332}
(1994), 36

\bibitem{super} N. Ohta and H. Tanaki, Phys. Rev. {\bf D44} (1991), 2607; A.
D'Adda, Class. Quant. Grav. {\bf 9} (1992), L21; H. J. Schnitzer, Mod. Phys.
Lett. {\bf A7} (1992), 2449

\bibitem{mz} Yu. Makeenko and K. Zarembo, Nucl. Phys. {\bf B422} (1994), 237;
J. Ambj\o rn, C. F. Kristjansen and Yu. Makeenko, Phys. Rev. {\bf D50} (1994),
5193; N. Marshall, G. W. Semenoff and R. J. Szabo, Phys. Lett. {\bf B351}
(1995), 153

\end{thebibliography}
\end{document}